\documentclass[11pt,twoside]{article}
\usepackage{asp2010}

\resetcounters

\begin{document}

\title{Discovery of a new Wolf-Rayet star using SAGE-LMC}
\author{V.V.~Gvaramadze$^1$, A.-N.~Chen\'{e}$^{2,3}$, A.Y.~Kniazev$^{4,1}$ and O.~Schnurr$^5$
\affil{$^1$Sternberg Astronomical Institute, Moscow State
University, Universitetskij Pr. 13, Moscow 119992, Russia}
\affil{$^2$Departamento de Astronom\'ia, Univ. de
Concepci\'on, Casilla 160-C, Chile} \affil{$^3$ Univ. de Valparaiso, Departamento de Fisica y Astronomia,
Avenida Gran Bretana 1111, Valparaiso, Chile, Chile} \affil{$^4$South African
Astronomical Observatory and Southern African Large Telescope
Foundation, PO Box 9, 7935 Observatory, Cape Town, South Africa}
\affil{$^5$AIP, An der Sternwarte
16, 14482 Potsdam, Germany}}

\begin{abstract}
We report the first-ever discovery of an extragalactic Wolf-Rayet
(WR) star with {\it Spitzer}. A new WR star in the Large
Magellanic Cloud (LMC) was revealed via detection of its
circumstellar shell using 24 $\mu$m images obtained in the
framework of the {\it Spitzer} Survey of the Large Magellanic
Cloud (SAGE-LMC). Subsequent spectroscopic observations with the
Gemini South resolved the central star in two components, one of
which is a WN3b+abs star, while the second one is a B0\,V star. We
consider the lopsided brightness distribution over the
circumstellar shell as an indication that the WR star is a runaway
and use this interpretation to identify a possible parent cluster
of the star.
\end{abstract}

\section{Introduction}

Massive stars are sources of copious stellar wind, which creates
circumstellar shells of a wide range of morphologies
\citep{no95,we01}. Detection of such nebulae by means of infrared
observations accompanied by spectroscopic follow-ups of their
central stars provide a useful tool for revealing evolved massive
stars \citep{gv09,gv10a,gv10b,gv10c,wa10,wa11}. In this paper, we report
the discovery of a new WR star in the LMC using this tool.

\section{A new WR star and its circular shell in the LMC}

\begin{figure}[!ht]
\includegraphics[width=13cm,angle=0,clip=]{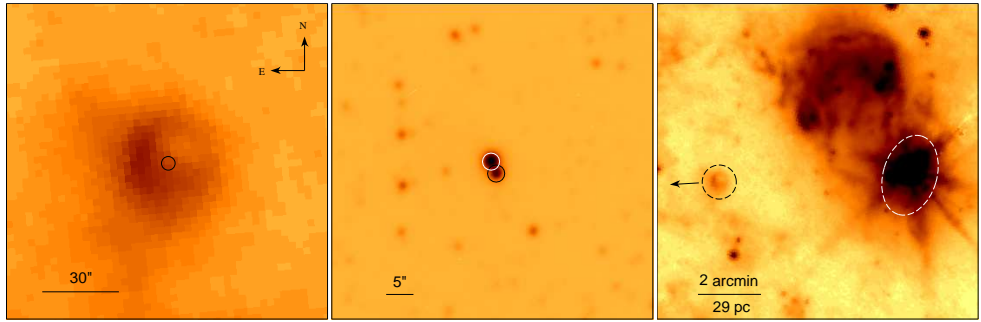}
\caption{{\it Left}: {\it Spitzer} 24\,$\mu$m image of the new
circular shell in the LMC. The position of the central star is
marked by a circle. {\it Middle}: Acquisition image of the central
star showing that it is composed of two components, star\,1 (new
WR star; indicated by a white circle) and star\,2 (black circle).
{\it Right}: 24\,$\mu$m image of the open cluster NGC 1722
(indicated by a dashed ellipse) with the position of the shell
marked by a circle. The arrow shows the direction of motion of the
WR star, as suggested by the brightness asymmetry of its shell.
The orientation of the images is the same.} \label{fig1}
\end{figure}

The new circular shell (Figure\,\ref{fig1}) was discovered using
the 24\,$\mu$m SAGE-LMC data \citep{me06} during our search for
bow shocks around runaway OB stars in the LMC \citep[for
motivation and the results of this search see ][]{gv10d}. To
determine the spectral type of the central star associated with
the shell we obtained its spectrum with the Gemini Multi-Object
Spectrograph South (GMOS-S). The acquisition image resolved the
central star in two components, hereafter star\,1 and star\,2 (see
middle panel of Figure\,\ref{fig1}).

\begin{figure}[!ht]
\includegraphics[width=9.5cm,angle=270,clip=]{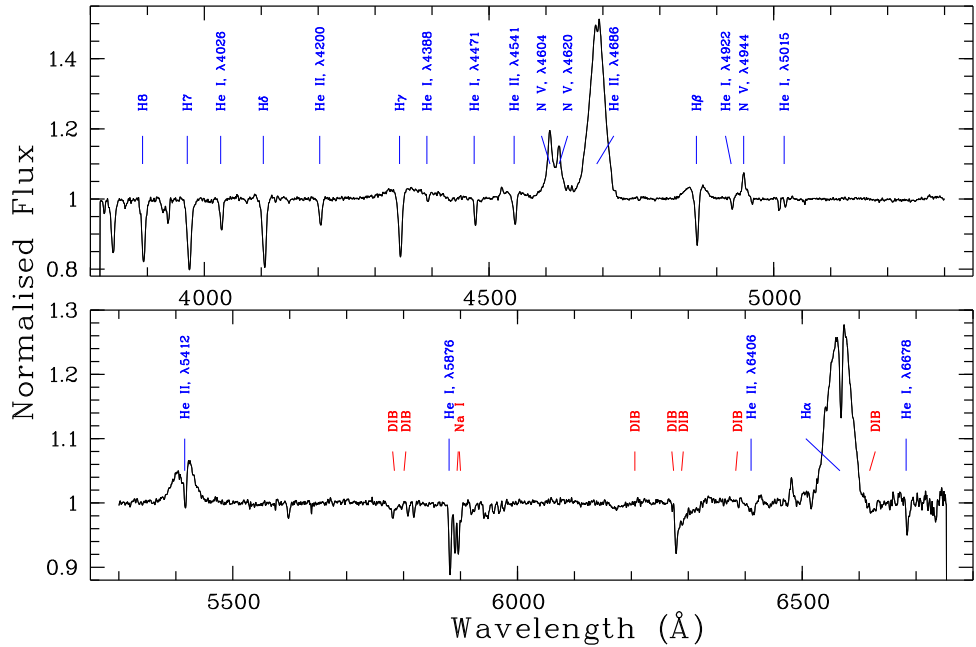}
\caption{Normalized spectrum of the new WR star (star\,1) in the
LMC with principal lines and most prominent diffuse interstellar
bands (DIBs) indicated.} \label{fig2}
\end{figure}

Figure\,\ref{fig2} shows the normalized spectrum of star\,1
obtained with the GMOS-S in a long-slit mode. The spectrum is
dominated by strong emissions of He\,{\sc ii} $\lambda$ 4686,
N\,{\sc v} $\lambda\lambda$ 4603, 4620, 4944 and H$\alpha$. The
H$\alpha$ and the He\,{\sc ii} $\lambda$ 4686 lines show central
absorption reversals. The spectrum also shows several
lower-intensity broad emission features with narrow absorption
reversals, of which the most prominent are H$\beta$ and He\,{\sc
ii} $\lambda 5412$. Other H and He lines are almost purely in
absorption. The presence of the N\,{\sc v} $\lambda\lambda$ 4603,
4620 emission lines and the absence of the N\,{\sc iii}
$\lambda\lambda$ 4634-41 multiplet implies that the emission
spectrum belongs to the ionization subclass WN3 \citep*{sm96}. The
FWHM of the He\,{\sc ii} $\lambda$ 4686 (emission) line of 31.1
\AA \, is quite large and star\,1 can be classified WN3b, but this
value is still very close to the empirically determined limit of
30 \AA \, for broad-lined stars \citep{sm96}. If star\,1 is really
a broad-line star \citep[which are believed to be the true H-free
WN stars;][]{sm98}, then the numerous absorption lines in its
spectrum might be caused by an unresolved companion massive star.
An indirect support to the binary status of star\,1 comes from its
position on the plot of EW(5411) versus FWHM(4686) for WN stars in
the LMC \citep[see Figure\,14 of ][]{sm96}, where it lies in a
region occupied by composite stars. Detailed comparison of the two
spectra of star\,1 collected on 2011 February 9 and 2011 March 5,
however, does not show any evidence of significant change in the
radial velocity. However, if the period is long and/or a multiple of our poor time sampling and/or if the eccentricity is
high, many more spectra are needed in order to really state if the star is a binary or not.
Alternatively, the absorption lines mimicking an O-type spectrum
could be intrinsic to star\,1 itself \citep[e.g.][]{fo03}.

The spectrum of star\,2 is dominated by H and He\,{\sc i}
absorption lines (Gvaramadze et al., in preparation). No He\,{\sc
ii} lines are visible in the spectrum. Using the
H$\gamma$-absolute magnitude calibration by \citet{ba74} and the
measured EW(H$\gamma$)=3.80 \AA, we estimated the spectral type of
star\,2 as B0\,V.

The lopsided brightness distribution over the shell and the offset
of the WR star from the geometric centre of the shell (see
Figure\,\ref{fig1}) suggest that the star is a runaway (moving
from west to east) and that the stellar wind interacts with dense
ambient medium (shed during the preceding evolutionary phase)
comoving with the star \citep{lo92,gv09}. Proceeding from this, we
searched for known star clusters to the west of the WR star and
found that this star was probably ejected from the open cluster
NGC 1722 (located at $\simeq 94$ pc in projection from the star;
see Figure \ref{fig1}).

\end{document}